\documentclass[aps,prl,twocolumn,superscriptaddress,letterpaper]{revtex4}

\usepackage{graphicx}
\usepackage{dcolumn}
\usepackage{bm}
\usepackage[latin1]{inputenc}
\newcommand{\tn}{\textnormal}

\begin{document}

\title{Violation of critical universality at the antiferromagnetic phase transition of YbRh$_2$Si$_2$}

\author{C. Krellner}
\email{krellner@cpfs.mpg.de}
\author{S. Hartmann}
\affiliation{Max Planck Institute for Chemical Physics of Solids, D-01187 Dresden, Germany }
\author{A. Pikul}
\affiliation{Max Planck Institute for Chemical Physics of Solids, D-01187 Dresden, Germany }
\affiliation{Polish Academy of Science, Institute of Low Temperature and Structure Research, 50-950 Wroclaw, Poland}
\author{N. Oeschler}
\author{J. G. Donath}
\author{C. Geibel}
\author{F. Steglich}
\affiliation{Max Planck Institute for Chemical Physics of Solids, D-01187 Dresden, Germany }
\author{J. Wosnitza}
\affiliation{Hochfeld-Magnetlabor Dresden (HLD), Forschungszentrum Dresden-Rossendorf, D-01314 Dresden, Germany}

\date{\today}

\begin{abstract}
We report on precise low-temperature specific-heat measurements, $C(T)$, of YbRh$_2$Si$_2$ in the vicinity of the antiferromagnetic phase transition on a single crystal of superior quality (RRR\,$\sim 150$). We observe a very sharp peak at $T_N = 72$\,mK with absolute values as high as $C/T=8$\,J/mol\,K$^2$. A detailed analysis of the critical exponent $\alpha$ around $T_N$ reveals $\alpha = 0.38$ which differs significantly from those of the conventional universality classes in the Ginzburg-Landau theory, where $\alpha \leq  0.11$. Thermal-expansion measurements corroborate this large positive critical exponent. These results provide insight into the nature of the critical magnetic fluctuations at a temperature-driven phase transition close to a quantum critical point.
\end{abstract}


\maketitle

Thermodynamically driven phase transitions (PTs) occur in manifold variants in solid-state physics. The Landau theory of PTs was a major achievement in describing PTs with one parameter, the order parameter \cite{Landau:1937}. Universal scaling dependences are valid for all continuous, i.e., second-order PTs and are only determined by the dimensionality of the system and the critical fluctuations \cite{Wosnitza:2007}. In contrast, quantum PTs are driven by quantum fluctuations associated with Heisenberg's uncertainty principle \cite{Sachdev:2001, Coleman:2005, Gegenwart:2008}. At a quantum critical point (QCP), the ordering temperature of a continuous classical PT is suppressed to zero temperature by a non-thermal control parameter (e.g., pressure or magnetic field), and the classical PT changes into a quantum one. Characteristics for a continuous quantum PT are strong deviations from Landau-Fermi-liquid theory in the physical properties of metallic systems in a broad range of the phase diagram around the QCP. In heavy-fermion metals, these non-Fermi-liquid phenomena have been studied with great care, but a complete theoretical concept replacing the Fermi-liquid paradigm is still lacking (for recent reviews see Ref.~\cite{Lohneysen:2007, Gegenwart:2008}). In this letter, we address the outstanding question whether a continuous magnetic PT driven by temperature in the vicinity of a QCP can be described within the framework of universality classes in the theory of PTs. To this end, detailed measurements of specific heat and thermal expansion of YbRh$_2$Si$_2$ around the antiferromagnetic (AFM) PT at $T_N=72$ mK were performed. 

The heavy-fermion system YbRh$_2$Si$_2$ is well suited to study this interplay between quantum and classical PTs, because it is a clean, stoichiometric, and well-characterized metal situated on the magnetic side ($T_N=72$ mK), but extremely close to an AFM QCP, leading to pronounced non-Fermi-liquid behavior in transport and thermodynamic properties, such as the divergence of the electronic Sommerfeld coefficient $\gamma=C^{4f}/T$, and a linear-in-T resistivity \cite{Trovarelli:2000}. The observed temperature dependences disagree with the expectation for the  three-dimensional (3D) spin-density-wave scenario that is applicable to many heavy-fermion materials at a QCP. Therefore, a new theoretical concept was developed, invoking critical excitations that are inherently quantum \cite{Si:2001, Coleman:2001}, explaining $\omega/T$ scaling in CeCu$_{5.9}$Au$_{0.1}$ as well as diverging Gr\"uneisen ratios \cite{Kuchler:2003, Tokiwa:2009} and a jump of the Fermi volume observed in Hall-effect measurements \cite{Paschen:2004} in YbRh$_2$Si$_2$. In this framework of `local' quantum criticality, the Kondo effect is critically destroyed because local moments are coupled not only to the conduction electrons but also to the fluctuations of the other local moments \cite{Gegenwart:2008}. The recent observation of an additional energy scale vanishing at the QCP which does neither correspond to the N\'eel temperature nor to the upper boundary of the Fermi-liquid region has boosted the interest in YbRh$_2$Si$_2$ \cite{Gegenwart:2007}.

High-quality single crystals of YbRh$_2$Si$_2$ were grown from indium flux. The growing parameters were improved to gain larger single crystals with higher crystallinity. This new generation of single crystals exhibits residual resistivities as low as $\rho_0\sim 0.5\,\mu\Omega$cm, corresponding to a residual resistivity ratio of $\sim 150$ \cite{Gegenwart:2007a}. The specific heat between 0.04 and 1\,K was measured in a $^3$He/$^4$He dilution refrigerator, utilizing a semi-adiabatic heat-pulse method with background heating ~\cite{Wilhelm:2004a, Radu:2005} on a platelet-like single crystal ($m\sim 10$\,mg) mounted on a silver platform. The addenda of the latter was measured separately and did not exceed $5\%$ of the total measured specific heat at temperatures below $1$\,K. The determination of the critical exponent necessitates a very stable and accurate temperature of the environment. We achieved a temperature noise level below 30\,$\mu$K and a resolution of the temperature measurement better than $\pm 5\,\mu$K. For temperatures above 0.5\,K, the specific heat was measured in a commercial PPMS with a $^3$He-insert. The 4$f$ contribution to the specific heat, $C^{4f}$, was obtained by subtracting the non-magnetic, $C_{Lu}$, and the nuclear, $C_Q$, contribution from the measured specific heat, $C_{meas}$, as $C^{4f}=C_{meas}-C_{Lu}-C_Q$. $C_{Lu}$ was determined by measuring the specific heat of the non-magnetic reference sample LuRh$_2$Si$_2$ below 10\,K \cite{Ferstl:2007}. However, $C_{Lu}$ at 1\,K contributes only to 1\,\% of $C_{meas}$ and the nuclear quadrupolar contribution, $C_Q=\alpha_Q/T^2$, calculated for Yb with $\alpha_Q=5.68\times 10^{-6}$\,JK/mol as determined from M\"ossbauer results \cite{Plessel:2003, Custers:2003}, is well below 1\,\% of $C_{meas}$ around $T_N$.

\begin{figure}[tb]
\includegraphics[width=8.5cm]{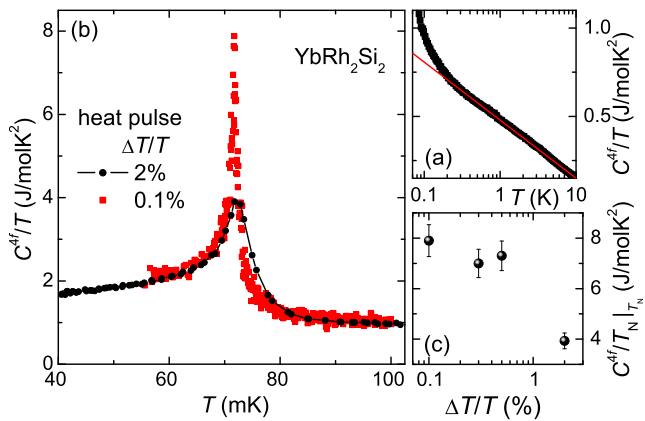}
\caption{\label{FigCTvsT} (Color online) (a) $4f$ increment to the specific heat plotted as  $C^{4f}/T(T)$ on a logarithmic $T$ scale for temperatures above $T_N$. Grey (red) line indicates a fit $C/T\propto \ln(T_0/T)$ with $T_0=30$\,K between 0.3 and 10\,K. (b) $C^{4f}/T(T)$ on a linear $T$-scale. The large peak at $T_N=72$\,mK shows the onset of AFM order. The peak sharpens dramatically upon reducing the power of the heat pulse during the measurement, leading from $\Delta T/T=2$\,\% (circles) to $0.1$\,\% (squares).  (c) Absolute values of $C^{4f}/T$ at $T_N$ as a function of $\Delta T/T$.} 
\end{figure}

In Fig.~\ref{FigCTvsT}a, the $4f$ increment to the specific heat is plotted as $C^{4f}/T$ on a logarithmic temperature scale above $T_N$. Below $T\sim 10$\,K, $C^{4f}/T$ logarithmically increases with $C^{4f}/T= a \ln (T_0/T)$ with $a=0.15$\, J/mol K$^2$ and $T_0=30$\,K, in agreement with results on samples of lower quality \cite{Trovarelli:2000, Gegenwart:2006}. Below $T\sim 0.3$\,K, $C^{4f}/T$ increases stronger than logarithmically with a power law of $C^{4f}/T\propto T^{-0.34}$. This additional upturn was discussed in detail in Ref.~\cite{Custers:2003} and ascribed to a `break-up' of the heavy quasiparticles at the QCP. We note that this unique upturn is also visible in our new higher-quality sample, in contrast to measurements reported by Knebel \textit{et al.}~\cite{Knebel:2006}.

Next, we focus on the specific heat at the magnetic PT. In Fig.~\ref{FigCTvsT}b, $C^{4f}/T$ is shown on a linear temperature scale. To determine the shape of the peak very accurately, we have reduced the power of the heat pulse to the lowest value at which an accurate determination of the heat capacity is still possible. In a standard experiment the power of the  heat pulse is chosen such that the achieved temperature increase, $\Delta T$, amounts to $\Delta T/ T\approx 2$\,\%. The latter value was found out to be a good compromise of large signal-to-noise ratio and reasonable resolution \cite{Radu:2005}. The measurement with $\Delta T/ T=2$\,\% yields the black curve in Fig.~\ref{FigCTvsT}b. A pronounced peak is observed at $T_N$. However, sharp features in the specific heat are artificially broadened when using such a relatively large $\Delta T/ T$. The measurements on the high-quality single crystal reveal a drastic sharpening of the peak when decreasing the power of the heat pulse such that $\Delta T/ T= 0.1$\,\%: The maximum value at $T_N$ reaches $C^{4f}/T\approx 8$\,J/mol K$^2$. In Fig.~\ref{FigCTvsT}c, this peak height is shown as function of $\Delta T/ T$. The absolute value of $C^{4f}/T$ at $T_N$ increases strongly for $\Delta T/ T< 2$\,\% but only slightly for $\Delta T/ T< 0.5$\,\%, in contrast to what one would expect for a first-order PT. The scattering of the data below and above $T_N$ is larger compared to the standard curve owing to the small temperature increase, $\Delta T\leq0.1$\,mK.

Well below $T_N$, the Sommerfeld coefficient becomes constant, $\gamma_0=1.7$\,J/mol K$^2$, in agreement with previous studies \cite{Gegenwart:2006}. However, the fact that $C^{4f}/T$ is larger below the magnetic PT than above is surprising and not yet understood. We note that the peak at $T_N$ is remarkably sharp while, usually in heavy-fermion systems, anomalies in the specific heat associated with the onset of magnetic order are broad \cite{Steglich:1996, Lohneysen:1996}. To the best of our knowledge there is no deep investigation of the critical exponent in specific-heat measurements on magnetically ordered heavy-fermion systems.

In order to classify the magnetic PT, a detailed analysis of the critical behavior was done. We applied the usual fit function \cite{Wosnitza:2007},
\begin{equation}\label{FktFit}
C^{\pm}(t) = \frac{A^{\pm}}{\alpha}|t|^{-\alpha} + b + Et \tn{ ,}
\end{equation}
to describe the critical behavior with the reduced temperature $t=(T-T_N)/T_N$; $+$($-$) refers to $t>0$ ($t<0$), respectively. The background contribution is approximated by a linear $t$ dependence ($b+Et$) close to $T_N$, while the power law (first term in Eq.~\ref{FktFit}) represents the leading contribution to the singularity in $C^{4f}(t)$ \cite{Kornblit:1973, Weber:1996}. Often, a higher-order correction term is introduced in the fitting function; however, as the leading contribution in the case of YbRh$_2$Si$_2$  is very large, the data can be described well with Eq. (\ref{FktFit}) \cite{Wosnitza:2007}. The procedure is as follows: The parameters $\alpha$, $A^{\pm}$, $b$, and $E$ are fitted with fixed $T_N=72$\,mK, simultaneously for $t>0$ and $t<0$. Then the fit is improved with an extended temperature range of typically $0.003\leq |t|\leq 0.1$. This routine is repeated for slightly different $T_N$ until the best fit (smallest deviation of the root-mean-square error) is obtained which, in addition, gives a very accurate value of $T_N$. We performed this fitting procedure for three experimental data sets (1, 2, 3) obtained with different $\Delta T/ T$ (0.5\%, 0.3\%, 0.1\%), the results of which are shown in Tab.~\ref{TabFit}. For fit No. 1a we used the same experimental data set as for No. 1 but before fitting we subtracted the specific heat obtained at the critical field ($T_N\rightarrow 0$), in order to account for the quantum fluctuations. The critical field within the easy plane amounts to $B_c=0.06$\,T and $C^{4f}/T$ grows as $T^{-0.34}$ below 0.3\,K at $B_c$ \cite{Oeschler:2008}. From Tab.~\ref{TabFit}, it is obvious that the different experimental data sets lead to almost the same critical exponent, $\alpha=0.38\pm0.03$. Moreover, $\alpha$ is not sensitive to the background subtraction which is completely different for fits 1 and 1a.

\begin{table}[tb]
\caption{\label{TabFit} Parameters derived from the fits of the specific-heat data for a high-quality single crystal of YbRh$_2$Si$_2$  around the AFM PT for different experimental data sets (see text). Fit 1a is based on the same experimental data as fit 1, but the contribution of the quantum fluctuations to the specific heat was subtracted prior to the fitting procedure. In fit 3, a Gaussian distribution of $T_N$ with $\delta T_N/T_N=3\cdot 10^{-4}$ was included.}
\begin{ruledtabular}
\begin{tabular}{ccccccccc}
Fit	&	$\Delta T/T$&	$T_N$	&	$A^{+}$  & $A^{+}/A^{-}$& $\alpha$ \\
	&	(\%)				&	(mK)	&	(mJ/molK)& 				&  				\\	\hline
1	& 	$0.5$				& 	71.3	&	15.4 	& 0.58			&	0.37		\\
1a	& 	$0.5$				& 	$71.3$	&	15.9	& 0.58			&	0.36 		\\
2	& 	$0.3$				& 	$72.0$	&	14.6	& 0.67			&	0.39 		\\		
3	& 	$0.1$				& 	$71.9$	&	14.7	& 0.68			&	0.39 		\\
\end{tabular}
\end{ruledtabular}
\end{table}

\begin{figure}[tb]
\includegraphics[width=8.5cm]{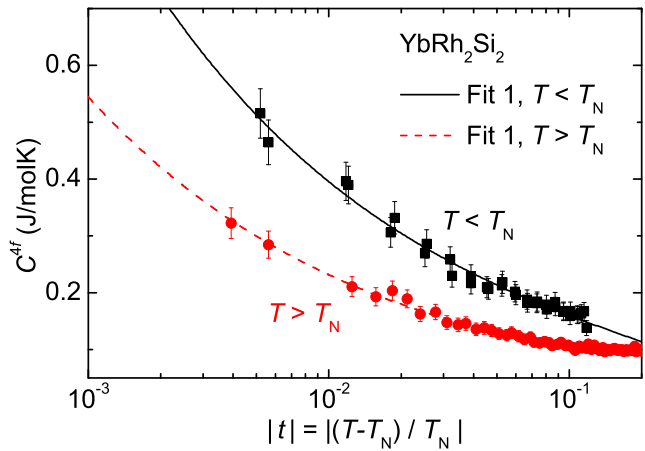}
 \caption{\label{FigCritExp} (Color online) Specific heat vs reduced temperature close to the AFM PT of YbRh$_2$Si$_2$. The data below and above $T_N$ can be best fitted with a critical exponent of $\alpha=0.38\pm0.03$.}
\end{figure}
The excellent quality of these fits becomes evident in Fig. \ref{FigCritExp}, where the curve for fit 1 is shown below (solid line) and above (dashed line) $T_N$, together with the experimental data (symbols). The total temperature window in Fig. \ref{FigCritExp} corresponds  to only 15\,mK below and above $T_N$. $C^{4f}(t)$ increases significantly close to $T_N$ which is a first sign for a large positive exponent $\alpha$. The fitting procedure described above yields a critical exponent of $\alpha=0.38\pm0.03$, which describes the data in the entire temperature range around the PT. Also, no saturation (due to rounding effects of the PT) is observed at the lowest  $|t|$ values. Only for $\Delta T/ T=0.1$\,\% we were able to measure below $|t|=10^{-3}$ where the data points tend to saturate. It may  be attributed to residual sample inhomogeneities producing a microscopic smearing of $T_N$ \cite{Kreps:1977}. We took these small rounding effects into account by a Gaussian distribution of $T_N$ ($\delta T_N/T_N=3\cdot 10^{-4}$) which has been carried out after the fitting procedure to Eq. (\ref{FktFit}) as described above. This method has been successfully applied to describe the rounding of specific-heat peaks \cite{Weber:1995, Deutschmann:1992}. The very narrow distribution $\delta T_N/T_N$ verifies the high quality of the YbRh$_2$Si$_2$  single crystal. The value of the critical parameter $\alpha$ is not sensitive to this additional parameter in fit~3, as can be seen from Tab.\,\ref{TabFit}.

\begin{table}[tb]
\caption{\label{TabCritExp}Theoretically predicted critical exponent $\alpha$ of the specific heat for different universality classes (after Ref.\,\cite{Wosnitza:2007}).}
\begin{ruledtabular}
\begin{tabular}{lcr}
Theory&$\alpha$\\
\hline
Landau mean field & 0 \\
2D-Ising & 0 \\
$n=1$, 3D-Ising & $+0.110(1)$ \\
$n=2$, 3D-$XY$ & $-0.015(1)$ \\
$n=3$, 3D-Heisenberg & $-0.133(5)$ \\ \hline
$n=2$, chiral-3D-$XY$ & $+0.34(6)$ \\
$n=3$, chiral-3D-Heisenberg  & $+0.24(8)$ \\
\end{tabular}
\end{ruledtabular}
\end{table}

In Tab.\,\ref{TabCritExp} we have summarized the theoretical values of $\alpha$ expected for the different universality classes, depending on the symmetry of the system. It is obvious that the observed exponent strongly deviates from all possible theoretical ones: The largest value for $\alpha$ is calculated for a 3D-Ising system, so that generally holds $\alpha\leq 0.11$. 

\begin{figure}[tb]
\includegraphics[width=8.5cm]{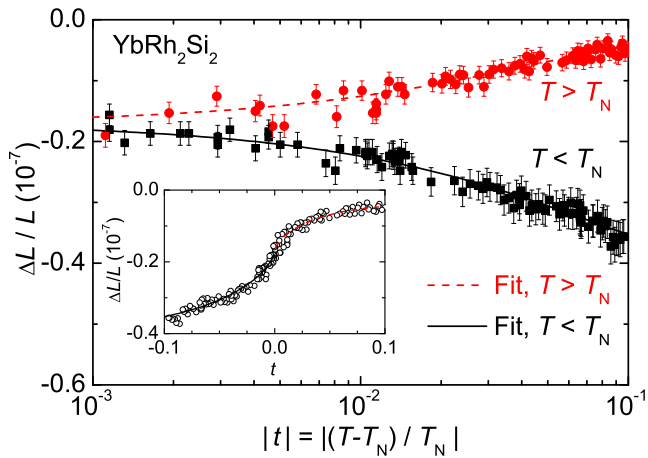}
 \caption{\label{FigCritExpdL} (Color online) Relative length change of YbRh$_2$Si$_2$ as function of the reduced temperature close to the AFM PT. The data below and above $T_N$ can be best fitted with a critical exponent of $\alpha'=-0.7\pm0.15$ which results in a critical exponent of the thermal expansion of $\alpha=0.3\pm0.15$ \cite{dL}. The inset presents the same set of data on a linear $t$-scale.}
\end{figure}

To establish the anomalous critical exponent we have performed thermal-expansion measurements in the vicinity of $T_N$. The length change was registered perpendicular to the tetragonal $c$ axis on the same single crystal used for the specific-heat measurements, described above. For details of the measurement setup, cf. Ref.~\cite{Kuchler:2003}. Thermodynamic relations reveal the same critical exponent for the thermal expansion as for the specific heat \cite{Scheer:1992}. However, it is more accurate to extract the critical exponent by fitting a modified Eq.~\ref{FktFit} directly to the measured relative length change, $\Delta L/L$, vs $t$ \cite{dL}. Application of the same fitting procedure as for the specific heat gives a critical exponent of the thermal expansion of $\alpha=0.30\pm0.15$, in good agreement with the specific-heat result. In Fig.~\ref{FigCritExpdL} this fit is shown below (solid line) and above (dashed line) $T_N$, together with the experimental data (symbols). However, the minor variation of $\Delta L /L$ around $T_N$ (inset of Fig.~\ref{FigCritExpdL}) severely complicates the analysis of the critical exponent and impedes a more accurate determination of $\alpha$. The absolute temperature calibration is not as accurate as for the specific-heat measurements which leads to a slightly lower $T_N=68.5$\,mK. This, however, does not influence the analysis of the critical exponent, performed in the same way as in the case of the specific-heat data. Therefore, our thermal-expansion results do confirm the surprisingly large critical exponent in YbRh$_2$Si$_2$.

The anomalous critical exponent $\alpha=0.38$ in YbRh$_2$Si$_2$ raises the question if this critical behavior might be indicating a tendency towards weak first-order behavior. However, no first-order PT in YbRh$_2$Si$_2$ is reflected either in the imaginary part of susceptibility measurements on this new generation of single crystals, or by hysteresis in any transport and thermodynamic property. In addition, no latent heat was resolved in our specific-heat measurements. Assuming a (somewhat smeared) first-order PT and integrating $C^{4f}$ over a temperature window of $| t | \leq 0.1$ one obtains $H\approx  1$\,mJ/mol as an upper boundary of the latent heat which indeed would be difficult to resolve.

Discarding of first-order PT, the value of $\alpha=0.38$ leads to the conjecture that the AFM order of YbRh$_2$Si$_2$  might belong to the new universality class of an XY antiferromagnet on a 3D stacked-triangular lattice, with $\alpha=0.34\pm0.06$ \cite{Kawamura:1992}, as experimentally confirmed on CsMnBr$_3$ \cite{Deutschmann:1992}. However, for YbRh$_2$Si$_2$ such a scenario is more than unlikely, because this system has no tendency to chirality. The latter would require either a triangular arrangement of the magnetic ions or a non-centro-symmetric Yb site, allowing the Dzyaloshinsky-Moriya interaction to operate. As long as the magnetic-ordering vector of the magnetic phase and the dispersion relation of the magnetic excitations are unknown, this question cannot be answered finally. Alternatively, the PT at $T_N$ in YbRh$_2$Si$_2$ may well be influenced by the pronounced quantum fluctuations of the nearby \textit{local} QCP which may substantially influence the spatial fluctuations of the classical order parameter and, thus, lead to internal degrees of freedom of the order parameter, preventing a description within the framework of the known universality classes.

In conclusion, we have presented high-accuracy specific-heat measurements in the very close vicinity of the antiferromagnetic phase transition of YbRh$_2$Si$_2$  resulting in an extremely sharp peak at $T_N=(72\pm1)$\,mK. A detailed analysis of the critical parameters yields an anomalous critical exponent $\alpha=0.38\pm0.03$ which cannot be explained within the conventional universality classes. Thermal-expansion measurements confirm this violation of critical universality close to a continuous quantum phase transition. Therefore, the observed large critical exponent poses a considerable challenge for those theories describing the magnetic phase transition and its critical fluctuations in the vicinity of a QCP.

We acknowledge fruitful discussions with M. Brando, P. Gegenwart, S. Kirchner, R. K\"uchler, C. P\'epin, Q. Si, and F. Weickert. The Deutsche Forschungsgemeinschaft (SFB 463, Research Unit 960) is acknowledged for financial support.


\begin{thebibliography}{32}
\bibitem[{dL()}]{dL}
\bibinfo{note}{Eq.~\ref{FktFit} modifies for the measured length change to
  $\Delta L/L(t) = \frac{A'}{\alpha'}t^{-\alpha'} + b' + E't$ (here only for
  $t>0$). From this it follows that the critical exponent of the thermal
  expansion is $\alpha=\alpha'+1$, because the thermal expansion is given by
  the temperature derivative of the relative length change $\Delta L/L$.}
\end{thebibliography}
\end{document}